\documentclass[11pt]{article}
\usepackage{geometry,epsf,amsmath,amsfonts,cite}

\usepackage{parskip}
\usepackage{amssymb}
\usepackage{braket}
\usepackage{url}
\usepackage[all]{xy}
\usepackage{subfig}
\usepackage[normalem]{ulem}
\usepackage{rotating}
\usepackage{rotfloat}
\usepackage{array}
\numberwithin{equation}{section}
\numberwithin{table}{section}
\newcommand{\Tr}{\mathrm{Tr}}
\newcommand{\tr}[1]{\mathrm{Tr}_L \left( #1 q^{L_0} \right)}
\newcommand{\trV}[1]{\mathrm{Tr}_V \left( #1 q^{L_0} \right)}
\DeclareMathOperator{\sump}{\sum_{p\geq 1}}
\DeclareMathOperator{\sumk}{\sum_{k\geq 1}}
\newcommand{\blank}[1]{}%

\begin{document}
\begin{flushright}  {~} \\[-12mm]
{\sf KCL-MTH-13-06}
\end{flushright} 

\thispagestyle{empty}

\begin{center} \vskip 24mm
{\Large\bf Characters of the $W_3$ algebra}\\[10mm] 
{\large 
Nicholas J. Iles
and
G\'erard M.\ T.\ Watts
}
\\[5mm]
Dept.\ of Mathematics, King's College London,\\
Strand, London WC2R\;2LS, UK
\\[5mm]

\vskip 4mm
\end{center}

\begin{quote}{\bf Abstract}\\[1mm]
Traces of powers of the zero mode in the $W_3$ Algebra have recently
been found to be of interest, for example in relation to Black Hole
thermodynamics, and arise as the terms in an expansion of the full
characters of the algebra. We calculate the first few such powers in
two cases. Firstly, we find the traces in the 3-state Potts model by
using null vectors to derive modular differential equations for the
traces.  Secondly, we calculate the exact results for Verma module
representations. We compare our two methods with each other and the
result of brute-force diagonalisation for low levels and find complete
agreement.
\end{quote}

{\setcounter{tocdepth}{2}
\tableofcontents
}

\newpage
\section{Introduction}
\label{char calc}

W-algebras are generalisations of the Virasoro algebra which appear in
a number of different situations. 
They all include the Virasoro algebra as a sub-algebra and so for any
highest-weight representation $V$ one can define a {\em reduced} character
as
\begin{equation}
  \chi_V^{red.\phantom|}\!(q)
= \mathrm{Tr}_V^{\phantom|}\!\big(\,q^{L_0 - c/24}\,\big)
\;.
\end{equation}
However, a W-algebra will typically have a larger set of commuting
zero-modes for which one can try to extend the reduced character to a
full character (or simply, character). Here we will only consider the
$W_3$ algebra which has one extra commuting mode\footnote{As the
Virasoro algebra has an infinite set of mutually commuting modes
\{$L_0$, $\Lambda_0$, \dots\}, so the $W_3$ algebra has \{$L_0$,
$W_0$, $\left(LW\right)_0 + \textrm{corrections}$, \dots\}. Here we
are looking at zero modes of fundamental fields $L$ and $W$ only.} $W_0$ 
and for which one can consider the character
\begin{equation}
  \chi_V^{\phantom|}(q,y)
= \mathrm{Tr}_V^{\phantom|}\!\big(\,q^{L_0 - c/24}\,y^{W_0}\,\big)
\;.
\end{equation}

\noindent These full characters are 
of particular interest in holography - see \cite{GabGopReview} for a review of the duality between $W_N$ CFTs and AdS$_3$ gravity theories. In \cite{GutKraus} for example, we see
that the $\mathrm{AdS}_3/\mathrm{CFT}_2$ correspondence gives us a
relation between the character of a Virasoro algebra module on the CFT
side, and the partition function on the AdS side. Generalising the
character to the $W_3$ algebra may allow the correspondence to be
investigated for a wider range of gravity models, those with spin
greater than 2 (see \cite{AGKP} for a review). 

Traces over W-algebra zero modes were calculated in \cite{KrausPerl} for the specific cases of free bosons and free fermions. Similarly, in \cite{GHJ} the authors calculate
$\mathrm{Tr}^{\phantom|}\left(W_0^n\hat{q}^{L_0-c/24}\right)$ for $n=2,4,6$ in the
limits $q\to 0$ and large central charge. An exact calculation of the
generalised character can both verify and extend their results.

\subsection{Content of the paper}

The ultimate aim is to calculate the full character of both the Verma
module and irreducible representations,
\begin{equation}
  \mathrm{Tr}_V^{\phantom|}(q^{L_0-c/24} y^{W_0})
= \mathrm{Tr}_V^{\phantom|}(q^{L_0-c/24} e^{2\pi i z W_0})
\;.
\end{equation}
In this paper we calculate the first few terms in the expansion of
this second form for several classes of expressions.  To make the
results simpler, we shall drop the extra term $(-c/24)$ from the
character which is usually included to improve modular transformation
properties. It is a simple matter to reinstate it if necessary.

To summarise the content of the rest of the paper:

In section \ref{direct calc}, we obtain explicit results up to level 6
for any 
highest weight Verma module
for the traces $\trV{W_0}$ through $\trV{W_0^5}$ by direct calculation.

In section \ref{Potts model}, we obtain exact expressions for
$\tr{W_0^2}$, and a differential equation whose solutions are exact
expressions for $\tr{W_0}$ for the irreducible minimal representations $L$ at
$c=4/5$, i.e. those relevant to the 3-state Potts model.

In section \ref{sec:vermachar} we find an exact (i.e. to all levels)
expression for 
the traces over any highest weight Verma module,
$\trV{W_0}$ and $\trV{W_0^2}$.

We compare our results from
these three methods of computation in section \ref{sec:compare} and show that they agree.  
Finally, we conclude with some open questions and directions for
future research.

\section{The $W_3$ algebra and its representations}

\subsection{The $W_3$ algebra}

The $W_3$ algebra is an extension of the Virasoro algebra
introduced by A.~Zamolodchikov in \cite{Zam} in which the usual spin-2
operators $L_n$ are augmented by spin-3 operators $W_n$. We will use
Zamolodchikov's normalisation so that the commutation relations are
\begin{subequations}
\label{W3 comm relns}
\begin{align}
\label{LL comm reln}
   \left[L_m\,,L_n\right] 
&= (m-n)L_{m+n} +
   \frac{c}{12}m(m^2{-}1) \delta_{m+n,0} 
\;,
\\
   \left[L_m\,,W_n\right] 
&= (2m-n)W_{m+n} 
\;,
\\
   \left[W_m\,,W_n\right] 
&= (m-n)
   \left[\frac{1}{15}\left(m{+}n{+}3\right)\left(m{+}n{+}2\right) -
         \frac{1}{6}(m{+}2)(n{+}2) \right]L_{m+n} \nonumber \\ 
&  \qquad{}+\beta(m{-}n)\Lambda_{m+n} +
   \frac{c}{360}m(m^2{-}1)(m^2{-}4)\delta_{m+n,0}
\;,
\end{align}
\end{subequations}
where
\begin{equation}
\beta=\frac{16}{22+5c}.
\end{equation}
The operators $\Lambda_n$ are given by
\begin{equation}
\label{defn of lambda_n}
  \Lambda_n 
= \sum_{p\leq -2} L_pL_{n-p} + \sum_{p\geq -1} L_{n-p}L_p 
 - \frac{3}{10}\left(n+2\right)\left(n+3\right)L_n 
= \sum_{p=-\infty}^{\lfloor \left(n-1\right)/2\rfloor} L_pL_{n-p} + \sum_{p=\lceil n/2 \rceil}^\infty L_{n-p}L_p + \gamma(n)L_n
\;,
\end{equation}
with 
\begin{equation}
\gamma(n)  = \begin{cases}
-\frac{1}{20}(n^2-4) & \hbox{$n$ even}\\
-\frac{1}{20}(n^2-9) & \hbox{$n$ odd}\\
\end{cases}
\end{equation}
Their commutation relations with $L_m$ and $W_m$ are
\begin{align}
   \left[L_m\,,\Lambda_n\right] 
&= \left(3m-n\right)\Lambda_{m+n} +
   \frac{22+5c}{30}m\left(m^2-1\right)L_{m+n} 
\;,\\
   \left[W_m\,,\Lambda_n\right] 
&= \frac{n{+}3}{5}\left[3\left(m{+}n{+}3\right)\left(m{+}n{+}4\right)
     - 8\left(m{+}1\right)\left(m{+}2\right)\right]W_{m+n} \nonumber
   \\
&  \qquad{}+ \sum_{p\leq -2} (2m{-}4n{+}4p )L_pW_{m+n-p} 
   + \sum_{p\geq -1}(2m{-}4n {+}4p)W_{m+n-p}L_p
\;.
\end{align}

\subsection{Representations of the $W_3$ algebra}
\label{W3 reps}

As for the Virasoro algebra, 
the $W_3$ algebra can have highest weight representations.
Since we have two commuting zero modes, highest weight representations
have two labels: the $L_0$ eigenvalue $h$ 
and the $W_0$ eigenvalue $w$. As such, highest-weight states are
labelled $\ket{h,w}$.
One can consider a wide range of different highest weight
representations, but we will only consider Verma modules and
irreducible modules. 

The Verma module $V_{h,w;c}$ is generated by the action of
the negative modes $W_{-m}, L_{-m}$ on the highest weight state
$\ket{h,w}$. One can show that a basis is taken by deciding on a fixed
ordering of 
these negative modes - any ordering will do - but we will choose a
particularly useful ordering in section \ref{sec:vermachar} when we
calculate the full character of the Verma module.

The irreducible module with highest weight $\ket{h,w}$, $L_{h,w;c}$,
is given as the quotient of the Verma module with the same highest
weight state by its maximal submodule.  Detailed (and apparently
correct) conjectures for the structure of the Verma modules of the
$W_3$ algebra and their relation to the irreducible modules are given
in \cite{dvvd}.

The reduced characters of the Verma module representations are trivial
to find, 
\begin{equation}
\label{red V chars}
  \chi^{red.}_{V_{h,w;c}}(q)
= \mathrm{Tr}^{\phantom|}_{V_{h,w;c}}(q^{L_0 - c/24})
= q^{h-c/24}\prod_{n=1}^\infty \frac{1}{(1-q^n)^2}
\;; 
\end{equation}
the reduced characters of {\em any} irreducible representation
can then be calculated using the results in \cite{dvvd}. In many cases
the reduced characters have alternative or closed-form expressions,
for example for the minimal models for which the characters are given
in \cite{minimal-model-characters}. We now summarise the results of
\cite{dvvd} that will be useful to us.

For any module $M$, one can form a \emph{local composition series}, a
sequence of submodules wherein each is itself a submodule of the
preceding one. The number of times a given module $M'$ appears in the
sequence is called the multiplicity of $M'$ in $M$. The multiplicity
is denoted $[M:M']$ and is independent of the particular sequence
chosen. For the particular case of $M=V$ a Verma module and $M'=L$ an
irreducible module, the multiplicities $[V:L]$ are invertible and we
denote these inverse multiplicities $(L:V)$. The \emph{Kazhdan-Lusztig
  conjecture} says that $[V:L]$ (resp. $(L:V)$) are given by the
Kazhdan-Lusztig polynomials (resp. inverse KL polynomials). 

In order to capture all contributions to $\chi_M$ without
over-counting, we can sum over characters of irreducible modules,
weighted by their multiplicities in $M$: $\chi_M = \sum_L [M:L]
\chi_L$, which for our purposes becomes $\chi_L = \sum_V (L:V)
\chi_V$. The inverse KL polynomials and the sum over Verma modules for
$W_N$ minimal models (such as the 3-state Potts model considered in
section \ref{Potts model}) are very simple: noting that $W_3$ is
associated to the Lie algebra $a_2$, the Verma modules that contribute
are those in representations related by the action of the affine Weyl
group of $a_2$ to the particular $L$ representation of interest, and
the inverse KL polynomials are $\pm 1$, where the sign depends on the
affine Weyl group element (as explained below). 

To make this concrete, let us introduce a new notation. Instead of
$h,w$, we can label representations by two pairs of integers
$mn;m'n'$, where each pair of integers is a weight of $a_2$, and a
parameter $t$ that labels the minimal model $M_{pp'}$ by
$t=p/p'$. This notation can label both Verma modules $V_{mn;m'n'}$ and
irreducible modules $L_{mn;m'n'}$ - for the latter, we have $m+n<p$,
$m'+n'<p'$, and $m,n,m',n'>0$. These new parameters are related to
$h$, $w$ and $c$ by \cite{BW92}
\begin{align}
  h\left(mn;m'n';t\right) 
&= \frac{1}{3t}\left[\left(m-m't\right)^2 +
   \left(m-m't\right)\left(n-n't\right) + \left(n-n't\right)^2 -
   3\left(1-t\right)^2\right]\\ 
\label{w mn defn}
w\left(mn;m'n';t\right) 
  &= \frac{\sqrt{2}}{9t\sqrt{3(5t{-}3)(5{-}3t)}}
  \big[  m{-} n {-} ( m'{-} n')t \big] 
  \big[ 2m{+} n {-} (2m'{+} n')t \big]
  \big[  m{+}2n {-} ( m'{+}2n')t \big]\\ 
c\left(t\right) &= 50 - 24t - \frac{24}{t} = 2\left(1-\frac{12\left(p-p'\right)^2}{pp'}\right)
\end{align}
\blank{
These weights are invariant under the action of the finite Weyl group of $a_2$ as follows:
\begin{equation}
\left[mn;m'n'\right] \equiv \left[ \left(p-m-n\right)m ;
  \left(p'-m'-n'\right)m'\right] \equiv \left[n\left(p-m-n\right) ;
  n'\left(p'-m'-n'\right)\right]. 
\end{equation}
}
The affine Weyl group is the semi-direct product of the finite Weyl
group with translations in the root lattice (the simple roots of $a_2$
are $\alpha_1 = (2,-1)$ and $\alpha_2 = (-1,2)$), and for $l(\omega)$ the
length of a finite Weyl group element $\omega$, the inverse KL polynomials
are simply $(-1)^{l(\omega)}$. Writing $\chi^{L}_{mn;m'n'} =
\mathrm{Tr}^{\phantom|}_{L_{mn;m'n'}}\left(Zq^{L_0}\right)$ for compactness, and
similarly for $\chi^V_{mn;m'n'}$, we therefore find 
\begin{align}
\label{alt sum}
\chi^L_{mn;m'n'} 
&= \sum_{r,s=-\infty}^\infty \sum_\omega (-1)^{l(\omega)} \chi^V_{\omega(m,n) + rp\alpha_1 + sp\alpha_2 ; m'n'}\\
&= \sum_{r,s=-\infty}^\infty \Big( 
  \chi^V_{m + 2rp - sp, n + 2sp -rp;m'n'} \;-\;
   \chi^V_{-m+2rp-sp,m+n+2sp-rp;m'n'}
\nonumber \\[-2mm]
&\qquad \qquad {}-\chi^V_{m+n + 2rp-sp,-n+2sp-rp;m'n'} + \chi^V_{n+2rp-sp, -m-n+2sp-rp;m'n'} \nonumber \\[1mm]
&\qquad \qquad  {}+ \chi^V_{-m-n+2rp-sp,m+2sp-rp;m'n'} - \chi^V_{-n+2rp-sp, -m+2sp-rp;m'n'} \;\;\Big).
\end{align}

For representations with $c>2$ there are only a finite number of null states and so these sums are even simpler. For example, the vacuum representation $\left[1,1;1,1\right]$ has character
\begin{align}
\label{vac irred}
\chi^L_{h=0,w=0,c} &= \chi^V_{1,1;1,1} - \chi^V_{-1,2;1,1} - \chi^V_{2,-1;1,1} + \chi^V_{1,-2;1,1} + \chi^V_{-2,1;1,1} - \chi^V_{-1,-1;1,1}  \\
&= \chi^V_{h=0,w=0,c} - \chi^V_{h=1,w=w_-,c} - \chi^V_{h=1,w=-w_-,c} + \chi^V_{h=3,w=w_+,c}  + \chi^V_{h=3,w=-w_+,c} - \chi^V_{h=4,w=0,c}
\end{align}
where (\ref{w mn defn}) gives
\begin{equation}
w_\pm = \frac{3\sqrt{2}\left(1\pm t\right)}{\sqrt{3\left(5t-3\right)\left(5-3t\right)}}.
\end{equation}


\section{Series-expansion solutions by direct calculation}
\label{direct calc}

Of course, once the commutation relations (\ref{W3 comm relns}) are
known, it is a simple (in principle) matter to ``brute force''
calculate $\trV{Z}$ or $\tr{Z}$ for any operator $Z$. 

For each state $\ket{S_l}$ in $V$ or $L$, where $l$ denotes the level,
commute $Z$ to the right (through the creation operators for that
state), and then commute the annihilation operators within $Z$ to the
right until they annihilate the highest-weight state $\ket{h,w}$. All
that remains will be $q^{h + l}$ times some linear combination of all
the states at level $l$. So, the contribution of $\ket{S_l}$ to the
trace is just $\alpha_S q^{h+l}$, where $\alpha_S$ is the coefficient
of $\ket{S_l}$ in the linear combination. 

Clearly though, this procedure will not give the complete answer as
there are infinitely many states in $V$ or $L$. However, 
we can use it level-by-level to obtain a series expansion in $q$. This
is more easily done for a Verma module $V$ since for an irreducible
module $L$, there is the additional complication of first deducing
which states are present (i.e. by finding the null states and removing
them). 

Using this method, we found the results given in 
Table~\ref{brute force results}, i.e.  
\begin{align}
\label{tr W0 expansion}
   \trV{W_0} 
&=   wq^h + 2wq^{h+1} + 5wq^{h+2} + 10wq^{h+3} + 20wq^{h+4}
   + 36wq^{h+5} + 65wq^{h+6} 
+ \ldots 
\\
   \trV{W_0^2} 
&= w^2q^h + \left(2w^2 +\frac{4}{22+5c}\left(32h-c+2\right)\right)q^{h+1} 
  + \dots \\
    \trV{W_0^3} 
&= w^3q^h + \left(2w^3 +\frac{12w}{22+5c}\left(32h-c+2\right)\right)q^{h+1}
 + \dots
\end{align}
etc.

\begin{table}
\renewcommand{\arraystretch}{1.2}
\centering
\raggedright
\begin{tabular}{c || c 
  | >{\raggedright}p{3.4cm} 
  | >{\raggedright}p{3.8cm} 
  | >{\raggedright}p{6.2cm} }
$n$ & 
$0$ & 1 & 2 & 3 
\tabularnewline \hline\hline

$W_0$ & 
$w$ & 
$2w$ & 
$5w$ & 
$10w$ 
\tabularnewline \hline

$W_0^2$ & 
$w^2$ & 
$2w^2$\\$ + \frac 14 \beta (32h{-}c{+}2)$ & 
$5w^2$\\$ + \beta(64h{-}c{+}18)$ & 
$10w^2$\\$ + \frac 14 \beta(928h{+}3c{+}634)$ 
\tabularnewline \hline

$W_0^3$ & 
$w^3$ & 
$2w^3$\\$ + \frac 34 w \beta (32h{-}c{+}2)$ & 
$5w^3 $\\$ + 3 w \beta (64h{-}c{+}18)$ & 
$10w^3 $\\$ + \frac 34 w \beta (928h{+}3c{+}634)$ 
\tabularnewline \hline

$W_0^4$ & 
$w^4$ & 
$2w^4 
$\\$+ \frac 32 w^2\beta (32h{-}c{+}2) 
$\\$+ \frac 1{32}\beta^2(32h{-}c{+}2)^2$ & 
$5w^4
$\\$+ 6 w^2 \beta  (64h{-}c{+}18) 
$\\$+ \frac 12 \beta^2
\big((64h{-}c{+}18)^2
  $\\~~~~~~~~$-64h(32h{-}c{+}2)
  $\\~~~~~~~~$-128(4h{+}1)\big)$ & 
$10w^4 
$\\$+ \frac32 w^2 \beta(928h{+}3c{+}634) 
$\\$+ \frac1{32}\beta^2(250880h^2 {+} 1728ch{+}512640h
  $\\~~~~~~~~${+}309c^2{-}5076c{+}127700)$ 
\tabularnewline \hline

$W_0^5$ & 
$w^5$ & 
$2w^5 
$\\$+ \frac52 w^3\beta(32h{-}c{+}2) 
$\\$+ \frac5{32} w \beta^2(32h{-}c{+}2)^2$ & 
$5w^5 
$\\$+ 10 w^3\beta(64h{-}c{+}18) 
$\\$+ \frac 52 w \beta^2
 \big((64h{-}c{+}18)^2
  $\\~~~~~~~~$-64h(32h{-}c{+}2)
  $\\~~~~~~~~$-128(4h{-}5)\big)$ & 
$10w^5 
$\\$+ \frac 52 w^3 \beta(928h{+}3c{+}634)
$\\$ + \frac 5{32} w \beta^2 (250880h^2 {+} 1728ch{+}512640h
  $\\~~~~~~~~${+}309c^2{-}5076c{+}447188 )$ 
\end{tabular}

\vspace{\baselineskip}

\begin{tabular}{c || >{\raggedright}p{6.8cm} | >{\raggedright}p{8.cm} }
$n$ & 
4 &
5 \tabularnewline \hline\hline

$W_0$ & 
$20w$ &
$36w$ \tabularnewline \hline

$W_0^2$ & 
$20w^2 + 16\beta(46h{+}c{+}49)$ &
$36w^2 + \frac 12 \beta(3616h{+}155c{+}5770)$ \tabularnewline \hline

$W_0^3$ & 
$20w^3 + 48 w \beta (46h{+}c{+}49)$ &
$36w^3 + \frac 32 w \beta (3616h{+}155c{+}5770)$ \tabularnewline \hline

$W_0^4$ & 
$20w^4 + 96 w^2\beta(46h{+}c{+}49)
$\\$\phantom{20 w^4} + 2 \beta^2(21248h^2 {+} 928ch {+}
66112h 
$\\~~~~~~~~~~~~~~~${+}73c^2 {+} 420c {+} 28340)$ &
$36w^4 + 3 w^2\beta(3616h {+} 155c {+} 5770) 
$\\$\phantom{36 w^4} +
\frac1{16}\beta^2(2450432h^2{{+}}212672ch
$\\~~~~~~~${+}11408000h{+}19681c^2{+}269820c{+}7354500)$  \tabularnewline \hline

$W_0^5$ & 
$20w^5 + 160 w^3 \beta(46h{+}c{+}49) 
$\\$\phantom{20 w^5} + 10 w \beta^2(21248h^2 {+} 928ch {+} 66112h 
$\\~~~~~~~~~~~~~~~${+}73c^2 {+} 420c {+} 79988)$ &
$36w^5 + 5 w^3\beta(3616h {+} 155c {+} 5770) 
$\\$\phantom{20 w^5}+ \frac5{16}w\beta^2(2450432h^2{+}212672ch
$\\~~~~~~~${+}11408000h{+}19681c^2{+}269820c{+}17492100)$ 
\end{tabular}

\vspace{\baselineskip}

\begin{tabular}{c || >{\raggedright}p{13.8cm} }
$n$ & 
6 \tabularnewline \hline\hline

$W_0$ & 
$65w$ \tabularnewline \hline

$W_0^2$ & 
$65w^2 + \beta(4384h{+}273c{+}8894)$ \tabularnewline \hline

$W_0^3$ & 
$65w^3 + 3w\beta(4384h{+}273c{+}8894)$ \tabularnewline \hline

$W_0^4$ & 
$65w^4 + 6w^2\beta(4384h{+}273c{+}8894) $\\$\phantom{65 w^4} + \frac 12 \beta^2(1037312h^2{+}129600ch{+}5983104h{+}14351c^2{+}254660c{+}5266492)$ \tabularnewline \hline

$W_0^5$ &
$65w^5 + 10w^3\beta(4384h{+}273c{+}8894) $\\$\phantom{65 w^4} + \frac 52 w \beta^2(1037312h^2{+}129600ch{+}5983104h{+}14351c^2{+}254660c{+}11256892)$ 
\end{tabular}

\vspace{\baselineskip}

\caption{The results of the brute-force calculation of $\Tr_{V_n}(Z)$ described in Section~\ref{direct calc} for the subspaces
  $V_n$ of the Verma module at level $n$ and for the insertions
  $Z=W_0,\dots,W_0^5$, up to $O\left(q^6\right)$} 
\label{brute force results}
\end{table}

\section{Exact results for irreducible modules in the 3-state Potts model}
\label{Potts model}

In Section \ref{direct calc}, we found series-expansion expressions for $\trV{W_0^n}$ up to $O(q^6)$. In this section, we use a different method to find an exact solution for $\tr{W_0}$ and $\tr{W_0^2}$. This method has the benefits of a) being exact, and b) giving a result for the irreducible module $L$, but the drawback of requiring one to choose a value of $c$. Despite this caveat, the method itself is applicable to any minimal model (i.e. any choice of $c$) for which one can find null vectors, and can be used to calculate $\tr{W_0^n}$ for any value of $n$.

For this method, we use the fact that\footnote{We use the field-state correspondence $\ket{F}=\lim_{z\to 0}F(z)\ket{0}$ and the Laurent expansion of the field $F(z)=\sum_m F_m z^{-m-h_F}$, where $h_F$ is the weight of $F$, to relate states, fields and operators/modes.}
\begin{equation}
\label{null field reln}
\tr{N_0}=0 \qquad \textrm{for any null field $N$.}
\end{equation}
This is immediately apparent from the definition of $L$, a module which has had all null states removed. The method is demonstrated below; here we simply give an explanation.

For every term in $\tr{N_0}$, we can use the commutation relations
(\ref{W3 comm relns}) and the cyclicity of the trace to move one of
the operators ($L_p$ or $W_p$, say) in that term all the way through
the other operators and back to its starting position. In doing so, we
pick up an overall factor of $q^p$ from moving the operator through
$q^{L_0}$, and introduce new terms from the commutation relations. If
these newly-introduced terms are comprised entirely of $L_0$ or $W_0$
operators, we are finished - if not, we repeat this procedure on the
remaining non-$L_0/W_0$ parts until only $L_0$ or $W_0$ remain. We
then rearrange the result to get an expression for our original term
solely in terms of traces of $L_0$ and $W_0$ operators, with some
coefficients that can depend on $q$, $p$ and $c$. 

Since $L_0$ can be considered to be a differential operator via
\begin{equation}
\label{L0 as diff op}
L_0q^{L_0}\ket{h,w} = L_0q^h\ket{h,w} = hq^h\ket{h,w} = q\frac{\partial}{\partial q}q^h\ket{h,w} = q\frac{\partial}{\partial q}q^{L_0}\ket{h,w},
\end{equation}
this can have one of two effects:
\begin{itemize}
\item if we have no mixed $\tr{L_0^mW_0^n}$ terms, it will turn every
  term in $\tr{N_0}$ into some differential operator acting on the
  reduced character $\tr{}$ - except for terms like $\tr{W_0^n}$. Since
  $\tr{N_0}=0$, we can thus write $\tr{W_0^n}$ in terms of known
  quantities: differential operators acting on the reduced character. 
\item if we do have mixed $\tr{L_0^mW_0^n}$ terms, we instead get
  differential operators acting on $\tr{W_0^n}$, and then $\tr{N_0}=0$
  gives a differential equation for $\tr{W_0^n}$ which we can solve to
  find $\tr{W_0^n}$. 
\end{itemize}
The first effect appears when we find $\tr{W_0^2}$, the second when we find $\tr{W_0}$.

In this section, we choose to work in the 3-state Potts model, as this has null fields at low levels, making calculations especially tractable. This is the $W_3$ minimal model $M_{5,4}$, i.e. $t=5/4$ and $c=4/5$, and it has the representations given in Table \ref{Potts reps}:
\begin{table}[h]
\renewcommand{\arraystretch}{2.1}
\[
\begin{array}{c||c|c|c|c|c|c}
h,w & 0,0 & \frac{1}{15},-\frac{1}{9}\sqrt{\frac{2}{195}} & \frac{1}{15},+\frac{1}{9}\sqrt{\frac{2}{195}} & \frac{2}{3},-\frac{2}{9}\sqrt{\frac{26}{15}} & \frac{2}{3},+\frac{2}{9}\sqrt{\frac{26}{15}} & \frac{2}{5},0 \\ \hline
mn;m'n' & 11;11 & 12;11 & 21;11 & 13;11 & 31;11 & 22;11 \\
\end{array}
\]
\caption{$W_3$ representations of the 3-state Potts model} 
\label{Potts reps}
\end{table}

\subsection{The trace $\tr{W_0}$}
For our calculation of $\tr{W_0}$, we will need a null field that
includes a term that is just $W(z)$ or its derivatives - such a field
can be found at level seven in the Potts model: 
\begin{equation}
N^7 = -\frac{1}{12}W'''' + \frac{52}{121}\left(W''L\right) +
\frac{47}{121} \left(W'L'\right) - \frac{27}{121}\left(W\Lambda\right)
+ \frac{141}{605}\left(WL''\right) 
\end{equation}
where the brackets $\left(\dots\right)$ denote normal ordering of the
enclosed fields by \cite{DMS}
\begin{equation}
\left(AB\right)_m = \sum_{n\leq -h_A} A_n B_{m-n} + \sum_{n> -h_A} B_{m-n} A_n.
\end{equation}
This field has zero mode 
\begin{align}
N^7_0 &= \sump \left(\frac{636}{605}p^2 + \frac{54}{121}p +
  \frac{126}{605}\right)L_{-p}W_p + \sump\left(\frac{636}{605}p^2 -
  \frac{54}{121}p + \frac{126}{605}\right)W_{-p}L_p \nonumber \\ 
\label{N70}
& \qquad {}- \frac{27}{121}\sump W_{-p}\Lambda_p - \frac{27}{121}\sump
\Lambda_{-p}W_p - \frac{27}{121}\Lambda_0W_0 + \frac{126}{605}L_0W_0 -
\frac{6}{605}W_0 
\end{align}
which we can then substitute into the null field relation (\ref{null field reln}).

Let us demonstrate the method outlined above in detail for the first term in (\ref{N70}):
\begin{align}
\tr{L_{-p}W_p} &= q^p \tr{W_pL_{-p}} \\
&= q^p \tr{L_{-p}W_p} + q^p \tr{\left[W_p,L_{-p}\right]} \\
&= \frac{q^p}{1-q^p}\tr{3pW_0}
\end{align}
giving a final expression for the trace over this first term of
\begin{gather}
\tr{\sump \left(\frac{636}{605}p^2 + \frac{54}{121}p +
    \frac{126}{605}\right)L_{-p}W_p} = \sump \frac{3pq^p}{1-q^p}
\left(\frac{636}{605}p^2 + \frac{54}{121}p + \frac{126}{605}\right)
\tr{W_0} \nonumber \\ 
= \left( \frac{1908}{605}\sump \frac{p^3q^p}{1-q^p} +
  \frac{162}{121}\sump \frac{p^2q^p}{1-q^p} + \frac{378}{605}\sump
  \frac{pq^p}{1-q^p} \right)\tr{W_0}. 
\end{gather}

Repeating this process for the remaining terms in (\ref{N70}), we end up with
\begin{equation}
f_2\tr{L_0^2 W_0} + f_1\tr{L_0W_0} + f_0 \tr{W_0} = 0,
\end{equation}
which gives (using (\ref{L0 as diff op}) and writing $\mathcal{W}\equiv\tr{W_0}$)
\begin{equation}
\label{W0 diff eqn}
q^2 f_2 \mathcal{W}'' + q\left( f_2 + f_1 \right) \mathcal{W}' + f_0 \mathcal{W} = 0.
\end{equation}
In these expressions, the coefficient functions $f_i$ are given by
\begin{align}
f_0 &= -\frac{1944}{121}\left(\sump \frac{pq^p}{1-q^p}\right)^{\!\!2}
- \frac{324}{121}\sump\frac{p^2q^p}{\left(1-q^p\right)^2} +
\frac{306}{55}\sump \frac{p^3 q^p}{1-q^p} +
\frac{882}{605}\sump\frac{pq^p}{1-q^p} - \frac{6}{605} \\ 
f_1 &= -\frac{432}{121}\sump\frac{pq^p}{1-q^p} + \frac{9}{55}, \\
f_2 &= -\frac{27}{121}.
\end{align}
These functions can also be expressed in terms of Eisenstein series
$E_{2k}(q)$ as
\begin{align}
f_0 &=
-\frac{3}{12100}(1 + 20 E_2 + 75 (E_2)^2 - 56 E_4 )
\;,\;\;\;\;
f_1 = \frac{9}{605}(1 + 10 E_2)
\;,
\end{align}
where here, and later, we use
\begin{equation}
E_{2} = 1 - 24 \sum_{n=1}^\infty \frac{n q^n}{1-q^n}\;,\;\;
E_{4} = 1 + 240 \sum_{n=1}^\infty \frac{n^3 q^n}{1-q^n}\;,\;\;
E_{6} = 1 - 504 \sum_{n=1}^\infty \frac{n^5 q^n}{1-q^n}\;.\;\;
\end{equation}

In principle, we now solve (\ref{W0 diff eqn}) to find
$\tr{W_0}$. 
However, it is clear that this is not a simple ODE to
solve exactly, but it does allow us to solve for a series expansion to
any desired order. 
The leading power of $q$ is determined by the indicial equation and
can only take the values $1/15$ or $2/3$. The series expansions can then
be found order-by-order.
The first few terms are 
\begin{subequations}
\label{W0 expr}
\begin{align}
\label{W0 expr 1/15}
\tr{W_0} &= \pm\frac{1}{9}\sqrt{\frac{2}{195}}q^\frac{1}{15}\left( 1 + 46q + 74q^2 + 192q^3 - 121q^4 + 286q^5 -314q^6 - 166q^7 + O\left(q^8\right) \right) \\
\intertext{and}
\label{W0 expr 2/3}
\tr{W_0} &= \pm \frac{2}{9}\frac{1}{\sqrt{15 \cdot 26}}q^\frac{2}{3}\left( 26 + 143q + 142q^2 + 214q^3 - 22q^4 - 23q^5 - 386q^6 + 26q^7 + O\left(q^8\right) \right).
\end{align}
\end{subequations}
We would like to point out that these are the unique non-trivial
series expansion solutions to the differential equation (\ref{W0 diff eqn}). Particularly, the leading powers of $q$, $1/15$ or $2/3$, are exactly the values of $h$ in the Potts model representations that have non-zero $w$ (see Table \ref{Potts reps}).

\subsection{The trace $\tr{W_0^2}$}
To calculate $\tr{W_0^2}$, we will need a null field that includes a
term like $(WW)(z)$, which we find at level six in the Potts model: 
\begin{equation}
N^6 = \left( WW \right) -\frac{85}{78} \left( L'L' \right) -
\frac{95}{117} \left( L\Lambda \right) - \frac{25}{234} L'''' +
\frac{29}{39} \left( LL'' \right), 
\end{equation}
which has zero mode
\begin{align}
N^6_0 &= 2\sump W_{-p}W_p - \frac{95}{117}\sump L_{-p}\Lambda_p -
\frac{95}{117} \sump \Lambda_{-p}L_p + \sump \left( \frac{11}{3}p^2 +
  \frac{8}{39}\right) L_{-p}L_p \nonumber \\ 
&\qquad\qquad\qquad {}+ W_0^2 - \frac{95}{117}\Lambda_0L_0 +
\frac{4}{39}L_0^2 - \frac{22}{195}L_0 + \frac{4}{9}\Lambda_0. 
\end{align}
Again, we substitute this into the null field relation (\ref{null
  field reln}) and perform the manipulations above to find 
\begin{align}
\tr{W_0^2} &= g_3 \tr{L_0^3} + g_2 \tr{L_0^2} + g_1 \tr{L_0} + g_0 \tr{} \\
\label{W0^2 diff eqn}
&= \left[ q^3 g_3 \frac{\partial^3}{\partial q^3} + q^2\left(3g_3 +
    g_2 \right) \frac{\partial^2}{\partial q^2} + q\left(g_3 + g_2 +
    g_1 \right) \frac{\partial}{\partial q} + g_0 \right] \tr{} 
\end{align}
with
\begin{align}
g_0 &=
\frac{1}{1263600} (-38 - 426 E_2 - 705 (E_2)^2 + 663 E_4 + 
   72 E_2 E_4 + 434 E_6)
\\
g_1 &=
\frac{1}{14040}(38 + 284 E_2 + 235 (E_2)^2 - 221 E_4)
\\
g_2 &= 
-\frac{1}{234}(19 + 71 E_2)
\;,\;\;\;\;\;\;
g_3 = \frac{95}{117}
\end{align}

As can be seen, this does not give a differential equation for
$\tr{W_0^2}$, but instead gives $\tr{W_0^2}$ in terms of a
differential operator acting on the reduced character $\tr{}$.

The reduced characters for the irreducible modules, $\tr{}$, are found
by applying (\ref{alt sum}) to the reduced Verma module characters
(\ref{red V chars}). So, since the coefficient functions $g_i$ and the
characters $\tr{}$ are known, we can evaluate the right-hand side of
(\ref{W0^2 diff eqn}) to find an expression for $\tr{W_0^2}$. The
solutions, which are composed of products of infinite sums, are not
particularly enlightening and so we do not state them here. Instead,
we give their series expansions up to $O(q^6)$ in table \ref{exact W0^2 for Potts}.
\begin{table}[h]
\renewcommand{\arraystretch}{1.9}
\[
\begin{array}{c | l}
h, w & q^{-h}\tr{W_0^2}\textrm{ to }O\left(q^6\right)\\  \hline\hline 
0, 0 & 12q^3 + \frac{4352}{65}q^4 + \frac{3064}{13}q^5 + \frac{50864}{65}q^6 \\  \hline 
\frac{1}{15}, \pm\frac{1}{9}\sqrt{\frac{2}{195}} & \frac{2}{15795} +
\frac{4232}{15795}q + \frac{22868}{3159}q^2 + \frac{227216}{5265}q^3 +
\frac{2965402}{15795}q^4 + \frac{9096296}{15795}q^5 +
\frac{23727188}{15795}q^6 \\ \hline  
\frac{2}{3}, \pm\frac{2}{9}\sqrt{\frac{26}{15}} & \frac{104}{1215} +
\frac{3146}{1215}q + \frac{365224}{15795}q^2 + \frac{279764}{3159}q^3
+ \frac{944152}{3159}q^4 + \frac{11270746}{15795}q^5 +
\frac{5613800}{3159}q^6 \\  \hline  
\frac{2}{5}, 0 & \frac{28}{13}q + \frac{256}{13}q^2 +
\frac{6872}{65}q^3 + \frac{20992}{65}q^4 + \frac{61872}{65}q^5
+\frac{140592}{65}q^6 \\  
\end{array}
\]
\caption{Series expansions for $\tr{W_0^2}$ in the three-state Potts
  model obtained from the exact expressions (not shown). Note that for the $h=1/15$ and $h=2/3$ representations, the expansions are the same for either choice of $w$ sign.} 
\label{exact W0^2 for Potts}
\end{table}

\newpage
\section{Exact results for Verma modules}
\label{sec:vermachar}
\label{proposed expressions}

The Verma module has a particularly simple and explicit basis which
means we can, at some effort, calculate the trace over a Verma module
of any particular power $W_0^n q^{L_0}$ by explicitly calculating the
action of $W_0^n$ at each level.
We shall take as our canonical basis states of the form
\begin{equation}
 \prod_{n=1}^\infty (\, L_{-n}^{a_n}\,W_{-n}^{b_n}\,) \ket{h,w}
= \cdots W_{-p-1}^{b_{p+1}} \,L_{-p}^{a_p} W_{-p}^{b_p}\,L_{-p+1}^{a_{p-1}} 
  \cdots \ket{h,w}
  \;,\;\; a_n,b_n \geq 0 \;.
\label{eq:basis}
\end{equation}

There are two simple results we need which will prove very useful. 
The first is that we can grade the monomials in the $W_3$-algebra by the
number of $W$-modes and by the total number of modes.
If a monomial has $w$ $W$--modes and $n$ modes in total, we say it has
grade $(w,n)$, and we can assign it a ``total grade'' $(w+n)$

The second is that the effect of commuting two modes changes the grade by
one of four possibilities, which are given in Table \ref{tab:grades}.
Since each of the commutators decreases the total grade, we are
see that the result of repeated commutators is guaranteed to terminate.
Secondly, we only need consider a finite (and small) number of
possible commutation relations between modes in a monomial to
calculate the contribution to the trace.

\blank{
\begin{table}
\renewcommand{\arraystretch}{1.4}
\[
\begin{array}{c|cc}
\hline
\hbox{commutator} & 
\hbox{mode} & 
\hbox{change in grade} \\
\hline
~[L,L] & \hbox{central} & (\,0,-2) \\
       & L              & (\,0,-1) \\
\hline
~[L,W] & W              & (\,0,-1) \\
\hline
~[W,W] & \hbox{central} & (-2,-2) \\
       & L              & (-2,-1) \\
       & LL             & (-2,\,0)  \\
\hline
\end{array}
\]
\caption{Possible changes in grade after a commutator.}
\label{tab:grades}
\end{table}
}

\begin{table}[htb]
\renewcommand{\arraystretch}{1.4}
\[
\begin{array}{c|cc|c|ccc}
\hbox{commutator} &
\multicolumn{2}{c|}{ ~[L,L] }&
\multicolumn{1}{c|}{ ~~~~~[L,W]~~~~ }&
\multicolumn{3}{c}{ ~[W,W] }
\\ \hline
\hbox{mode} &
\hbox{central} & L &
 W &
\hbox{central} & L & LL 
\\ \hline
\hbox{change in grade} 
& (\,0,-2)  & (\,0,-1)  & (\,0,-1)  & (-2,-2)  & (-2,-1)  & (-2,\,0)  
\end{array}
\]
\caption{Possible changes in grade after a commutator.}
\label{tab:grades}
\end{table}

\subsection{The trace $\mathrm{Tr}_V(q^{L_0})$}

We first restate the result for the reduced character of the Verma
module,
\begin{equation}
  \mathrm{Tr}^{\phantom|}_{V_{h,w;c}}(q^{L_0})
= q^h\,\prod_{n=1}^\infty \frac{1}{(1-q^n)^2}
= \frac{q^h}{\phi(q)^2}
\;.
\end{equation}
The two factors $(1-q^n)$ in the denominator correspond to the (equal)
contributions from the modes $W_{-n}$ and $L_{-n}$. 

\subsection{The trace $\mathrm{Tr}_V(W_0 \,q^{L_0})$}

The next simplest result is for the trace with the insertion of a
single $W_0$ mode. 
Let us denote a state in the basis (\ref{eq:basis}) as ${\cal P}
\ket{h,w}$ where ${\cal P}$ is a monomial in the lowering modes of the
algebra. We have
\begin{equation}
 W_0 {\cal P}\ket{h,w}
= [W_0,{\cal P}] \ket{h,w} + w {\cal P} \ket{h,w}
\;.
\end{equation}
Suppose ${\cal P}$ has W-number $w$. After the action of a single
commutator and no further re-arranging, $[W_0,{\cal P}]$ has terms
with W-number $(w+1)$ 
arising from the commutator $[W_0,L_{-p}]=2p W_{-p}$ (for some $p$)
and terms with W-number $(w-1)$ arising from the commutator
$[W_0,W_{-p}]$ for some $p$. Since W-number
is non-increasing, the terms with W-number $(w-1)$ cannot contribute
to the trace, so we can restrict attention to the terms where
$[W_0,L_{-p}]=2p W_{-p}$.
These terms have W-number $w+1$, and so we need to reduce this by 1 if
we are to get a contribution to the trace. The only way to do that is
to create a mode $W_0$ during the rearrangement of the terms which
can act on the highest weight state, but such a mode can only be
created by the action $[L_p,W_{-p}]$ and the mode $L_p$ can itself
only come from the expansion of a term $\Lambda_{-q}$, which would
itself come from a W-number reducing commutator $[W_r,W_s]$. The
result is that such a term cannot arise and the only contribution to
the trace of $W_0$ comes from the action of the mode $W_0$ on the
highest weight state itself. This gives the immediate result
\begin{equation}
\label{W0 V expr}
  \mathrm{Tr}^{\phantom|}_{V_{h,w;c}}(\,W_0\,q^L_0)
= w\, q^h\,\prod_{n=1}^\infty \frac{1}{(1-q^n)^2}
= \frac{w\,q^h}{\phi(q)^2}
\;.
\end{equation}

\blank{
If we examine the action on our particular basis (\ref{eq:basis}), we
see that these terms are of the form
\begin{equation}
 W_0 \;(\cdots L_{-p}^a W_{-p}^b\cdots)\ket{h,w}
\longrightarrow
 (\cdots L_{-p}^{a'} W_{-p}^{\phantom a}\, L_{-p}^{a-a'-1} W_{-p}^b \cdots)\ket{h,w}
\;.
\end{equation}
The rearrangement of such a state into the canonical ordering will
only require the moving of $W_{-p}$ past $L_{-p}$, creating at most $W_{-2p}$
W-parity is not preserved under the action on the highest weight state
of course, since $w_0\ket{h,w} = w\ket{h,w}$ so the only way the
W-parity of $[W_0,{\cal P}]\ket{h,w}$ can be equal to that of 
${\cal P}\ket{h,w}$ is if an extra mode $W_0$ is created during the
rearrangement of the terms in $[W_0,{\cal P}]$ into the canonical
order in (\ref{eq:basis}).
We note that The commutator of $W_0$ with any term in the monomial ${\cal P}$ will
change the W-parity, since
}

This agrees with the result given in section \ref{direct calc}, see Table \ref{brute force results}
and (\ref{tr W0 expansion}), that 
\begin{align}
   \mathrm{Tr}^{\phantom|}_{V_{h,w;c}}( W_0 \, q^{L_0} )
&= wq^h + 2wq^{h+1} + 5wq^{h+2} + 10wq^{h+3} + 20wq^{h+4} 
   + 36wq^{h+5} + 65wq^{h+6} + \dots 
\end{align}

\subsection{The trace $\mathrm{Tr}_V(\,W_0^2\, q^{L_0})$}

We can calculate the trace of $W_0^2$ on one of the basis states
(\ref{eq:basis}) by keeping track of the possible
ways in which the action of $W_0^2$ can reproduce the original state.
We have
\begin{equation}
  W_0^2 {\cal P}\ket{h,w}
= w^2 {\cal P}\ket{h,w}
+ 2 w [W_0\,,{\cal P}]\,\ket{h,w}
+ [W_0\,,[W_0\,,{\cal P}]]\,\ket{h,w}
\label{eq:act2}
\end{equation}
The first term is simply the action of $W_0^2$ on the highest weight
state, giving a term in the character
\begin{equation}
\frac{ w^2\,q^h }{\phi(q)^2}
\;.
\label{eq:term1}
\end{equation}
The second term does not contribute to the trace, as we have already
considered exactly this term in the previous section.

The third term is more involved. We calculate this term by
splitting the action of the $W_0$--mode into several pieces, and by
considering the possible action on the modes in the monomial on a
case-by-case basis.

\newcommand{\ds}{\displaystyle}

It is helpful to split each $[W_0\,,W_n]$ commutator into various separate
terms and analyse the contributions from each of these terms to the
trace. We split the commutator up into five terms $(a)$--$(e)$ as

\begin{equation}
\renewcommand{\arraystretch}{1.4}
\begin{array}{rcl@{~~~~~}l}
  {}[W_0\,,W_{-m}]
&=& 
\ds
  \;\alpha_r \,L_{-m}
&(a)
\\
&&\ds + \, m\beta\,\epsilon_m \,L_{-m/2} \,L_{-m/2} &(b)\\
&&\ds + \; 2 m \beta \sum_{s=\lceil (m+1)/2 \rceil}^{m-1} L_{-s}L_{-m+s} &(c)\\
&&\ds + \; 2 m \beta \,L_{-m} L_0 &(d)\\
&&\ds + \; 2 m \beta \sum_{s=m+1}^{\infty} L_{-s}L_{-m+s} &(e)
\;,
\end{array}
\label{eq:wwsplit}
\end{equation}
where 
\begin{equation}
\alpha_r = \left( \frac{r(r^2-4)}{15} + r \beta\gamma(r) \right)
\;,\;\;\;\;
\epsilon_r = 1 \hbox{ if $r$ is even and 0 otherwise.}
\end{equation}


Next, we list the modes in a monomial which are acted
upon by the commutators $[W_0,\cdot\,]$. 
There are nine possibilities listed in Table \ref{tab:modes_hit}
labelled (1)--(9). The
arrows indicate which modes are acted upon by each $[W_0,\cdot\,]$; a
double arrow indicates the same mode is acted upon twice. We consider
these nine possibilities in turn.

\begin{table}[htb]
\renewcommand{\arraystretch}{1.3}
\[
\begin{array}{l|l|l|l|l}
(1) & (2) & (3) & (4) & (5) \\
\hline
 {\overset{\textstyle\downdownarrows}{L}}_{-r} & 
 {\overset{\textstyle\downdownarrows}{W}}_{-r} &
 {\overset{\textstyle\downarrow}{L}}_{-r} \cdots  {\overset{\textstyle\downarrow}{L}}_{-r} &
 {\overset{\textstyle\downarrow}{L}}_{-r} \cdots  {\overset{\textstyle\downarrow}{W}}_{-r} &
 {\overset{\textstyle\downarrow}{W}}_{-r} \cdots  {\overset{\textstyle\downarrow}{W}}_{-r} 
\\
\hline\hline
\multicolumn{2}{l|}{(6) } & (7) & (8) & (9) \\
\hline
\multicolumn{2}{l|}{ {\overset{\textstyle\downarrow}{L}}_{-r} \cdots  {\overset{\textstyle\downarrow}{L}}_{-s} }&
 {\overset{\textstyle\downarrow}{L}}_{-r} \cdots   {\overset{\textstyle\downarrow}{W}}_{-s} &
 {\overset{\textstyle\downarrow}{W}}_{-r} \cdots   {\overset{\textstyle\downarrow}{L}}_{-s} &
 {\overset{\textstyle\downarrow}{W}}_{-r} \cdots   {\overset{\textstyle\downarrow}{W}}_{-s} 
\end{array}
\]
\caption{possible choices for the action of $W_0$; note that $r>s$.}
\label{tab:modes_hit}
\end{table}

\subsubsection{Contributions from term (1)}

We consider all contributions from terms where $W_0$ acts twice on the
same mode $L_{-r}$. The first action gives
\newcommand{\be}{\begin{equation}}
\newcommand{\ee}{\end{equation}}
\be
~[W_0,L_{-r}] = 2r\, W_{-r}
\ee
The second gives five terms, from the five contributions in equation
(\ref{eq:wwsplit}).
We list these in turn.

\begin{itemize}
\setlength{\itemindent}{.7cm}

\item[(1)(a):]
$[W_0,[W_0,L_{-r}]] \to 2r\alpha_r\,L_{-r}$

This directly contributes to the trace. For the mode $L_{-r}$ we get a
term $2r\alpha_r q^r/(1 - q^r)^2$, and for all the other modes
$X_{-s}$ we get $1/(1-q^s)$, so that the total contribution is
\be
T_{1a}=
  \frac{q^h}{\phi(q)^2}\,\sum_{r=1}^\infty
  \frac{2r\,\alpha_r\,q^r}{1-q^r}
\;.
\ee

\item[(1)(b):]
$[W_0,[W_0,L_{-r}]] \to 2r^2\,\beta\,\epsilon_r\,L_{-r/2}\,L_{-r/2}$.

The action of the $W_0$ modes has removed a mode $L_{-r}$ from the
monomial and replaced it by two modes $L_{-r/2}$, so that the grade
has increased by $(0,1)$. 
To recover this lost mode, we can consider at most one commutator of
the form $[L,L]$ or $[L,W]$, which change the grade by $(0,-1)$, but
no such commutator will produce the desired mode, so this term does
not contribute to the trace,

\item[(1)(c):]
  $[W_0,[W_0,L_{-r}]] \to 4r^2\,\beta\sum_s L_{-s} L_{-r+s}$.

We need to reorder the monomial into canonical order, which we do by
first moving the mode $L_{-r+s}$ to the right, and then the mode
$L_{-s}$. In the process of the first move, we have to commute
$L_{-r+s}$ past possible modes $L_{-p}$ and modes $W_{-p}$. Only one
such commutator can reinstate the missing mode $L_{-r}$, which is 
$[L_{-r+s},L_{-s}] = (2s-r)L_{-r}$. The total contribution from all
such terms is thus
\be
T_{1c} = 
4\beta \frac{q^h}{\phi(q)^2}
\sum_{r=1}^\infty r^2\frac{q^r}{1-q^r}
\sum_{s>r/2}^{r-1} (2s-r) \frac{q^s}{1-q^s}
\ee

\item[(1)(d):]
$[W_0,[W_0,L_{-r}]] \to 4r^2\,\beta\,L_{-r}L_0$

This again contributes directly to the trace, since $L_0$ acts on all
the remaining terms in the monomial and the highest weight state as
well.
The action on the highest weight state gives the simple contribution
\be
T_{1di}=
4h\beta\frac{q^h}{\phi(q)^2}\,\sum_{r=1}^\infty \frac{r^2q^r}{1-q^r}
\ee
The action of $L_0$ on the remaining modes gives the more complicated
contribution
\begin{eqnarray}
T_{1dii}&=&
q^h
\sum_{r=1}^\infty 
4 r^2\beta
  \left[\prod_{k=r+1}^\infty\frac{1}{(1-q^k)^2}\right]
\frac{q^r}{1-q^r} 
  q\frac{d}{dq}\left[ \prod_{k=1}^r \frac{1}{(1-q^k)^2}\right]
\\
&=&
4\beta\frac{q^h}{\phi(q)^2}
\sum_{r=1}^\infty \frac{r^2\,q^r}{1-q^r} \sum_{k=1}^r \frac{2k\,q^k}{1-q^k}
\end{eqnarray}

\item[(1)(e):]
$[W_0,[W_0,L_{-r}] \to 2r\,\beta\sum_s L_{-s}L_{-r+s}$

This has removed a mode $L_{-r}$ from the monomial and (by considering
the grade) we are only allowed to use a single commutator to return
this to the monomial. This could only arise by moving $L_{-r+s}$ or $L_{-s}$
to their canonical positions, but neither of these could create
$L_{-r}$ in this manner and so this term does not contribution to the trace.

\end{itemize}

\subsubsection{Contributions from term (2)}

We next consider all contributions from terms where $W_0$ acts twice on the
same mode $W_{-r}$. The first action gives five terms, one for each of
the possible terms in (\ref{eq:wwsplit}). We consider the action of
$W_0$ on each of these five terms in turn

\begin{itemize}
\setlength{\itemindent}{.7cm}

\item[(2)(a):]
$[W_0,[W_0,W_{-r}]] \to [W_0, \alpha_rL_{-r}] = 2r\alpha_r\,W_{-r}$

This directly contributes to the trace, giving the same result as term
(1)(b):
\be
T_{2a}=
  \frac{q^h}{\phi(q)^2}\,\sum_{r=1}^\infty
  2r\,\alpha_r\frac{\,q^r}{1-q^r}
\;.
\ee

\item[(2)(b):]
$[W_0,[W_0,W_{-r}]] \to [W_0,r\,\beta\,\epsilon_r\,L_{-r/2}\,L_{-r/2}]
= r^2\,\beta\,\epsilon_r\,( W_{-r/2}L_{-r/2} + L_{-r/2}W_{-r/2} )$.

The action of the $W_0$ modes has removed a mode $W_{-r}$, and
inserted modes $W_{-r/2}$ and $L_{-r/2}$ which need to be commuted to
the right. In the process, one of the modes $W_{-r/2}$ can commute
with a mode $L_{-r/2}$ giving $[W_{-r/2},L_{-r/2}]=(r/2)W_{-r}$. 
The two terms $W_{-r/2}L_{-r/2}$ and $L_{-r/2}W_{-r/2}$ differ in
their contribution because the first term allows for this extra
commutator giving $W_{-r}$, and together they give
\be
T_{2b}=
\frac\beta 2 
\frac{q^h}{\phi(q)^2}\,
\sum_{r=2,\,\mathrm{even}}^\infty r^3\,
  \frac{1+q^{r/2}}{1-q^{r/2}}\,\frac{q^r}{(1-q^r)}
\;.
\ee

\item[(2)(c):]
$[W_0,[W_0,W_{-r}]] \to 
 \sum_s 2r\beta [W_0,L_{-s} L_{-r+s}]
= 2r\beta \sum_s( 2s W_{-s}L_{-r+s} + 2(r-s) L_{-s}W_{-r+s})
$

We consider the two terms separately. 
In the first term, we first move $L_{-r+s}$ to the right; doing so, it
can commute with a mode $W_{-s}$ to give $[L_{-r+s},W_{-s}] =
(3s-2r)W_{-r}$, restoring the mode $W_{-r}$. Moving the mode $W_{-s}$
to the right cannot give a mode $W_{-r}$ since $(-r+s)>-s$ and so
$W_{-s}$ will not pass through any modes $L_{-r+s}$.
Hence the first term gives
\be
T_{2ci}=
  4\beta\frac{q^h}{\phi(q)^2}
\sum_{r=1}^\infty \frac{r q^r}{1-q^r}
\sum_{s>r/2}^{r-1} \frac{ s(3s-2r) q^s}{1-q^s}
\;.
\ee
The second term contributes when $W_{-r+s}$ commutes through a mode
$L_{-s}$ giving $(3s-r)W_{-r}$, and a total contribution
\be
T_{2cii}=
  4\beta\frac{q^h}{\phi(q)^2}
\sum_{r=1}^\infty \frac{r q^r}{1-q^r}
\sum_{s>r/2}^{r-1} \frac{ (r-s)(3s-r) q^s}{1-q^s}
\;.
\ee

\item[(2)(d):]
$[W_0,[W_0,W_{-r}] \to 2r\,\beta\,[ W_0,L_{-r}L_0]
= 4r^2\beta W_{-r} L_0$

This again contributes directly to the trace, since $L_0$ acts on all
the remaining terms in the monomial and the highest weight state as
well.
The action on the highest weight state gives the simple contribution
\be
  T_{2di}
=
  4h\beta\frac{q^h}{\phi(q)^2}\,\sum_{r=1}^\infty \frac{r^2 q^r}{1-q^r}
\ee
The action of $L_0$ on the remaining modes gives the more complicated
contribution
\begin{eqnarray}
T_{2dii}&=&
\sum_{r=1}^\infty 
4 r^2\beta q^h
  \left[\prod_{k=r}^\infty\frac{1}{(1-q^k)^2}\right]
  q^r q\frac{d}{dq}\left[ \frac{1}{1-q^r} 
\prod_{k=1}^{r-1} \frac{1}{(1-q^k)^2}\right]
\\
&=&
4\beta\frac{q^h}{\phi(q)^2}
\sum_{r=1}^\infty \frac{r^2 q^r}{1-q^r} 
\left( \frac{r q^r}{1-q^r} + \sum_{k=1}^{r-1} \frac{2kq^k}{1-q^k} \right)
\\
&=&
4\beta\frac{q^h}{\phi(q)^2}
\sum_{r=1}^\infty \frac{r^2 q^r}{1-q^r} 
\left( \sum_{k=1}^{r} \frac{2kq^k}{1-q^k} - \frac{r q^r}{1-q^r}  \right)
\end{eqnarray}

\item[(2)(e):]
$[W_0,[W_0,W_{-r}] \to 2r\,\beta\,\sum_s [ W_0,L_{-s}L_{-r+s}]
= 2r\beta\sum_s ( 2s W_{-s}L_{-r+s} + 2(s-r)L_{-s}W_{-r+s})$

As with term (1)(e), this has extra modes, either $W_{-s}$ or
$L_{-s}$, which 
cannot be removed by any commutator, and so case (2)(e) does not
contribute to the trace.

\end{itemize}

\subsubsection{Contributions from term (3)}

In term (3), each $W_0$ commutes with a mode $L_{-r}$ giving
$4r^2 W_{-r}W_{-r}$. This differs by grade $(0,2)$ so we would
need to reduce the number of $W$--modes by 2 to get a contribution to
the trace. This can only happen if one or both of the surplus $W_{-r}$
modes commutes with another $W$--mode while moving into its canonical
ordering. However, it does not pass through any such modes and so no
commutators arise and this gives no contribution to the trace.

\subsubsection{Contributions from term (4)}

In term (4), there are again five contributions from $[W_0,W_{-r}]$
which we again deal with in turn. There is also an overall factor 2
since the $W_0$ modes can act on the two modes in either order.

\begin{itemize}
\setlength{\itemindent}{.7cm}

\item[(4)(a):]
$[W_0,L_{-r}]\cdots[W_0,W_{-r}]
\to 4r \alpha_r\, W_{-r}\cdots L_{-r}$

This has grade $(0,0)$ and directly contributes to the trace a term
\be
T_{4a}=
4\frac{q^h}{\phi(q)^2}
\sum_{r=1}^\infty
 r\alpha_r\,\frac{q^{2r}}{(1-q^r)^2}
\;.
\ee

\item[(4)(b):]
$[W_0,L_{-r}]\cdots[W_0,W_{-r}]
\to 4r^2\beta \,W_{-r}\cdots L_{-r/2}L_{-r/2}$

We need to recover the mode $L_{-r}$ but that is not possible from any
of the reorderings and so it gives no contribution to the trace.

\item[(4)(c):]
$[W_0,L_{-r}]\cdots[W_0,W_{-r}]
\to 8r^2\beta \sum_s\, W_{-r}\cdots L_{-s}L_{-r+s}$

The only way this contributes to the trace is from the commutator
$[L_{-r+s},L_{-s}]=(2s-r)L_{-r}$ when moving the mode $L_{-r+s}$ to
the right. This leads to the total term
\be
T_{4c}=
8\beta\frac{q^h}{\phi(q)^2}
\sum_{r=1}^\infty r^2 \frac{q^{2r}}{(1-q^r)^2}
\sum_{s>r/2}^{r-1} (2s-r) \frac{q^s}{1-q^s}
\ee

\item[(4)(d):]
$[W_0,L_{-r}]\cdots[W_0,W_{-r}]
\to 8r^2\beta \, W_{-r}\cdots L_{-r}L_{0}$

This contributes directly to the trace when the mode $L_0$ commutes
through the modes to the right of it, and from the highest weight
state.
The action on the highest weight state gives the simple contribution
\be
  T_{4di}
= 8h\beta\frac{q^h}{\phi(q)^2}\,\sum_{r=1}^\infty \frac{r^2 q^{2r}}{(1-q^r)^2}
\ee
The action of $L_0$ on the remaining modes gives the more complicated
contribution
\begin{eqnarray}
T_{4dii}&=&
\sum_{r=1}^\infty 
8 r^2\beta q^h
  \left[\prod_{k=r}^\infty\frac{1}{(1-q^k)^2}\right]
  \frac{q^{2r}}{1-q^r} q\frac{d}{dq}\left[ \frac{1}{1-q^r} 
\prod_{k=1}^{r-1} \frac{1}{(1-q^k)^2}\right]
\\
&=&
8\beta\frac{q^h}{\phi(q)^2}
\sum_{r=1}^\infty \frac{r^2 q^{2r}}{(1-q^r)^2} 
\left( \frac{r q^r}{1-q^r} + \sum_{k=1}^{r-1} \frac{2kq^k}{1-q^k} \right)
\\
&=&
8\beta\frac{q^h}{\phi(q)^2}
\sum_{r=1}^\infty \frac{r^2 q^{2r}}{(1-q^r)^2} 
\left( \sum_{k=1}^{r} \frac{2kq^k}{1-q^k} - \frac{r q^r}{1-q^r} \right)
\end{eqnarray}

\item[(4)(e):]
$[W_0,L_{-r}]\cdots[W_0,W_{-r}]
\to 8r^2\beta \sum_s\, W_{-r}\cdots L_{-s}L_{-r+s}$

There is again no way to eliminate the mode $L_{-s}$ with $s>r$ so
this does not contribute to the trace.

\end{itemize}

\subsubsection{Contributions from term (5)}

The commutator $[W_0,W_{-r}]\cdots[W_0,W_{-r}]$ reduces the $w$-number
of the monomial by two. Since the $w$-number is non-increasing under
commutators, this does not contribute to the trace.

\subsubsection{Contributions from terms (6)}

In term (3), each $W_0$ commutes with a mode of the Virasoro algebra
giving a change in mode of $(2,0)$:
$[W_0,L_{-r}]\cdots[W_0,L_{-s}] = 4rs\,W_{-r}\cdots W_{-s}$.
We need to reduce the $w$-number which needs a $[W,W]$ commutator, but
these do not arise when moving $W_{-r}$ and $W_{-s}$ into canonical
ordering, so there is no contribution to the trace.

\subsubsection{Contributions from terms (7)}

We have the commutators
$[W_0,L_{-r}]\cdots[W_0,W_{-s}]$ with $r\neq s$.
This changes the grade of the monomial by $(0,0)$ or $(0,1)$ allowing
at most one commutator.
We have removed the mode $W_{-s}$ from the monomial and added the mode
$W_{-r}$. We cannot remove the extra mode $W_{-r}$ when it is moved into
canonical position and so there is no contribution to the trace.

\subsubsection{Contributions from terms (8)}

We have the commutators
$[W_0,W_{-r}]\cdots[W_0,L_{-s}]$ with $r\neq s$.
We have again changed the grade of the monomial by $(0,0)$ or $(0,1)$
allowing again at most one commutator.
This time we have added a mode $W_{-s}$ and we can remove it if a mode
$L_p$ included in the $[W_0,W_{-r}]\to LL$ terms commutes with it when
moving to canonical position. We have at the same time removed a
$W_{-r}$ mode, which needs to be replaced. This is possible if
$p-s=-r$.
This can come from either the terms (b) or (d) in the $[W_0,W_{-r}]$
commutator.

\begin{itemize}
\setlength{\itemindent}{.7cm}

\item[(8)(b):]
$
[W_0,W_{-r}]\cdots[W_0,L_{-r/2}]
\to
(r\beta L_{-r/2}L_{-r/2})\cdots (r W_{-r/2})
\to 
r^3\beta\,L_{-r/2}\cdots W_{-r}
$

This gives a contribution
\be
T_{8b}=
\beta \frac{q^h}{\phi(q)^2}
\sum_{r=2,\, \mathrm{even}}^\infty r^3\frac{q^{2r}}{(1-q^r)(1-q^{r/2})^2}
\ee

\item[(8)(c):]
$
[W_0,W_{-r}]\cdots[W_0,L_{-s}]
\to
\sum_s (2r\beta L_{-s}L_{-r+s})\cdots (2s W_{-s})
\to 
4r \sum_s s(3s-2r)\beta\,L_{-s}\cdots W_{-r}
$

Remembering to include all the possible $L_{-r/2}$ modes in the
monomial as well as those created in $[W_0,W_{-r}]$, 
this gives a contribution
\be
T_{8c}=
8\beta \frac{q^h}{\phi(q)^2}
\sum_{r=1}^\infty r\frac{q^r}{1-q^r}
\sum_{s>r/2}^{r-1} s(3s-2r) \frac{q^s}{(1-q^s)^2}
\ee

\end{itemize}

\subsubsection{Contributions from terms (9)}

The change in mode from $[W_0,W_{-r}]\cdots[W_0,W_{-s}]$ is $(-2,0)$,
$(-2,1)$ or $(-2,2)$. Since no re-orderings can possibly increase the
number of $W$-modes, there are no contributions to the trace from these terms.

\subsubsection{Combining all contributions}

This gives the final expression for the trace over the Verma module as
\begin{eqnarray}
 \Tr_V^{\phantom|}(\,W_0^2\,q^{L_0}\,)
&=&
  T_{1a} + T_{1c} + T_{1di} + T_{1dii} + 
  T_{2a} + T_{2b} + T_{2ci} + T_{2cii} +
  T_{2di} + T_{2dii}
\nonumber\\&&  +\; T_{4a} + T_{4c} +
  T_{4di} + T_{4dii} + T_{8b} + T_{8c}
\end{eqnarray}

The final result can be simplified by combining some of the terms. 
The terms with a
factor $h$ simplify as
\be
  T_{1di} + T_{2di} + T_{4di}
= 8h\beta \frac{q^h}{\phi(q)^2} \sum_{r=1}^\infty \frac{r^2 q^r}{\left(1-q^r\right)^2}
\;.
\ee
The terms with $\alpha_r$ combine as
\be
 T_{1a} + T_{2a} + T_{4a}
= 4 \frac{q^h}{\phi(q)^2} \sum_{r=1}^\infty r\alpha_r \frac{q^r}{\left(1-q^r\right)^2}
\;.
\ee
The terms coming from the derivative with respect to $q$ combine as
\be
T_{1dii} + T_{2dii} + T_{4dii}
= 4\beta\frac{q^h}{\phi(q)^2}
  \sum_{r=1}^\infty \frac{r^2 q^r}{(1-q^r)^2}\left(
-rq^r\frac{1+q^r}{1-q^r} + \sum_{k=1}^r \frac{4kq^k}{1-q^k}\right)
\;.
\ee
The sums over $r$ even can be combined as
\begin{eqnarray}
T_{2b} + T_{8b}
&=& 
\frac{\beta}{2} \,\frac{q^h}{\phi(q)^2}\,
\sum_{r=2,\,\mathrm{even}}^\infty
r^3\frac{1+q^r}{1-q^r} \frac{q^r}{(1-q^{r/2})^2}
\\
&=& 
4\beta \,\frac{q^h}{\phi(q)^2}\,
\sum_{r=1}^\infty
r^3\frac{1+q^{2r}}{1-q^{2r}} \frac{q^{2r}}{(1-q^{r})^2}
\;.
\end{eqnarray}
These in turn can then be combined with the terms from the derivatives
to give
\be
T_{1dii} + T_{2dii} + T_{4dii} + T_{2b} + T_{8b}
= 4\beta\frac{q^h}{\phi(q)^2}
\sum_{r=1}^\infty \frac{r^2q^r}{(1-q^r)^2}
\left( -\frac{2r q^r}{1-q^{2r}} + \sum_{k=1}^r \frac{4kq^k}{1-q^k}\right)
\ee
Finally, four of the terms with sums over $r/s<s<r$ can be combined as
\be
T_{1c} + T_{2ci}+ T_{2cii} + T_{4c}
= 8\beta \frac{q^h}{\phi(q)^2}
  \sum_{r=1}^\infty r^2 \frac{q^r}{(1-q^r)^2} 
  \sum_{s>r/2}^{r-1} (2s-r)\frac{q^s}{1-q^s}
\;.
\ee
Putting this all together, we arrive at an expression for the trace
over the Verma module as
\be\label{W0^2 expr}\renewcommand{\arraystretch}{1.8}
\Tr_V^{\phantom!}(W_0^2\,q^{L_0}\,)
= \frac{q^h}{\phi(q)^2}\left[
\begin{array}{c}
\ds
w^2
\;\;+\;\; \frac{4}{15} \sum_{r=1}^\infty \frac{r^2(r^2-4)q^r}{(1-q^r)^2}
\\
\ds
+ \;4 \beta \sum_{r=1}^\infty \frac{r^2q^r}{(1-q^r)^2}
   \left[ \,2h +
   \gamma(r) - 2 \frac{r q^{2r}}{1-q^{2r}}
  + 4 \sum_{k=1}^r\frac{kq^k}{1-q^k}
\right]
\\
\ds
+ \;8\beta
\sum_{r=1}^\infty \frac{r q^r}{1-q^r}
\sum_{s>r/2}^{r-1}
\frac{q^s}{1-q^s}
\left[
   \frac{r(2s-r)}{1-q^r}
+ \frac{s(3s-2r)}{1-q^s} \right]
\end{array}\right]
\ee

Despite our best efforts, we have not as yet been able to write this
result in a neater form.
Applying (\ref{alt sum}) to this directly does not yield a tractable
result. 

\section{Comparison of results}
\label{sec:compare}

To compare the results from section \ref{sec:vermachar}
 to those in section \ref{Potts model}, we need to convert from a trace over the Verma module $V_{h,w;c}$ to a trace over the irreducible module $L_{h,w;c}$ using the methods described in section \ref{W3 reps}: by applying (\ref{alt sum}) to our expressions for $\trV{W_0}$, given in (\ref{W0 V expr}), and $\trV{W_0^2}$ from (\ref{W0^2 expr}), we can find expressions for $\tr{W_0}$ and $\tr{W_0^2}$ as desired. We must then series-expand these expressions for the case of the 3-state Potts model, as this is how our irreducible module results are presented in section \ref{Potts model}.

For $\tr{W_0}$ with $h=1/15$ and $h=2/3$\footnote{Note that as $w=0$ for $h=0$ or $h=2/5$, we have $\trV{W_0}=0$ and hence $\tr{W_0}=0$, which of course trivially satisfies (\ref{W0 diff eqn}), for these representations.} we find (\ref{W0 expr 1/15}) and (\ref{W0 expr 2/3}) respectively, and for $\tr{W_0^2}$, the results we obtain coincide exactly with those given in Table \ref{exact W0^2 for Potts}. The agreements persist as far as we have calculated, $O(q^{30})$.

In summary then, in section \ref{sec:vermachar} we found exact Verma module expressions at any value of $c$ for $\trV{W_0}$ and $\trV{W_0^2}$, and verified that they agreed with the ``brute force'' expansions from section \ref{direct calc}. We then used the techniques described in section \ref{W3 reps} to convert these into irreducible module expressions and, for the specific example of the Potts model, compared them at the level of series expansions to the results obtained in section \ref{Potts model}. Again, they agreed as far as we calculated. Due to the unwieldy form of (\ref{W0^2 expr}), and the complicated form of (\ref{W0^2 diff eqn}), we have not yet been able to show that applying (\ref{alt sum}) to (\ref{W0^2 expr}) is identically equivalent to applying the differential operator (\ref{W0^2 diff eqn}) to the exact expressions for the reduced characters, although it should in principle be possible. Likewise, we have not yet been able to show that the expressions for $\tr{W_0}$ obtained from (\ref{W0 diff eqn}) and from applying (\ref{alt sum}) to (\ref{W0 V expr}) are equivalent.

\section{Future research directions}
\label{future research}
We conclude our discussion with a mention of possible directions that
future related research may take. 

We 
think the results here are interesting in their own right, and can be
used in other areas but we would obviously like to extend the results
we have here, both to general minimal models and to traces of higher
powers of $W_0$. Expressions are known for singular vectors in
$W_3$-minimal models \cite{BW92} and it may be possible to use these
to find traces 
of powers of $W_0$ in those representations.


Equations for Virasoro characters have been found from particle-like descriptions for a variety of particle content. In the 3-state Potts model for example, results have been found that can be interpreted as one-particle bosonic, two-particle bosonic, fermionic with two quasi-particles, and fermionic with one quasi-particle and two ghost particles\cite{DKMM}. It would be interesting to know if results could similarly be found for $W_3$ characters in this way, and if so how they relate to the expressions given in this work.

Calculations of the partition
function for $ \mathrm{Tr}\left(q^{L_0}\right)$ and $
\mathrm{Tr}\left(y^Qq^{L_0}\right)$, where $Q$ is a $\mathrm{U}(1)$
charge, can be performed using the modular transformations of these
traces, for example in \cite{AGKP}. Determining the modular transformation properties of the
generalised $W_3$ characters $ \mathrm{Tr}\left(y^{W_0}q^{L_0}\right)$
may offer an alternative route to results for the character, or indeed
vice versa: our results may help to shed light on the transformation
properties.

Many of the interesting applications of the results presented in this paper rely on their modular transformation properties: for example, $\hat{q}=\exp(-2\pi i / \tau)$ in \cite{GHJ} below is given by the action of the $SL(2,\mathbb{Z})$ generator $\mathcal{S}:\tau\to -1/\tau$ on $q=\exp(2\pi i \tau)$. Unfortunately, these properties are not currently known in general, and na\"{i}ve application of $\mathcal{S}$ to our results does not yield useful expressions. The modular transformation properties may need to be deduced before the research directions below and in the following section become feasible, nevertheless we state them here as we believe them to be interesting.

In \cite{GHJ}, the authors calculate the quantities
$\mathrm{Tr}\left(W_0^n\hat{q}^{L_0-c/24}\right)$ for $n=2,4,6$ in the
limits $q\to 0$, i.e. $\hat{q} \to 1$, and large central charge, in the $W_\infty[\lambda]$
algebra. Using the relation
\begin{equation}
W_\infty[\lambda]|_{\lambda=N} \cong W_N
\end{equation}
from \cite{GabGopTri} (see below), we could take the appropriate limits of our results and test them against the
results in \cite{GHJ} with $\lambda=3$.

We can use (\ref{W0^2 expr}) to calculate the leading term in $\mathrm{Tr} \! \left(W_0^2 \hat{q}^{L_0 - c/24}\right)$ over a single representation, e.g. a Verma module or the vacuum representation (using (\ref{vac irred})), as $q\to 0$, and we do not reproduce the results of \cite{GHJ}. It is not clear at the moment whether this is due to the differences between considering the full spectrum of the theory and a single representation, the continuum representation content in the modular transform of a single representation for $c>2$, or yet other effects. This is currently the subject of study.

We can find differential equations for the $q$-dependence of $\mathrm{Tr}_L \! \left(W_0 \hat{q}^{L_0 - c/24}\right)$ and $\mathrm{Tr}_L \! \left(W_0^2 \hat{q}^{L_0 - c/24}\right)$ in the 3-state Potts model by applying a modular transformation to equations (\ref{W0 diff eqn}) and (\ref{W0^2 diff eqn}). Again, the analysis of the resulting equations is currently under study.

\subsection{Holography}

As mentioned above, the quantities we have investigated in this work
are of relevance to holography, see \cite{AGKP} for a review. Interest in AdS$_3$/CFT$_2$ holography was sparked by \cite{GabGopHolo}, in which the authors proposed, and gave evidence to support, a duality between higher-spin Vasiliev theories on AdS$_3$, which are based on the higher-spin algebras $\textrm{hs}\left[\lambda\right]$, and $W_N$ minimal models in the large $N$ limit. The symmetry algebra of these large-$N$ models is thought to be $W_\infty\left[\lambda\right]$, where the 't Hooft coupling $\lambda=N/\left(N+k\right)$ is kept fixed as $N$ and the level $k$ are taken to infinity\footnote{For the unitary minimal models, i.e. those with $p'=p+1$, we have $p=k+3$. The Potts model, for example, corresponds to $k=1$.}. Black holes in these bulk theories were found in \cite{KrausPerl} and checked against the dual CFT results for free fermions and free bosons ($\lambda=0$ and 1 respectively), and the agreement was extended to general $\lambda$ in \cite{GHJ}, as described above. We may relate $W_\infty\left[\lambda\right]$ results to $W_N$ quantities using the algebra isomorphisms \cite{GabGopTri}
\begin{equation}
W_{N,k} \cong W_\infty\left[\frac{N}{N+k}\right]  \cong W_\infty\left[N\right].
\end{equation}

As the works cited above, and others, extended the study of the CFT side of this duality from $\lambda=0,1$ to generic $\lambda$ at large $c$, so our work further extends the understanding of these theories to include information about the generalised character for all values of $c$. We hope this will be of interest in a range of holographic applications. For example, in \cite{GutKraus} the
authors discuss the relation between the generalised character and the
partition function $Z_{bh}$ to which a BTZ black hole, generalised to carry spin-3 charge, contributes:
\begin{equation}
\label{holo reln}
Z_{bh}\left(\tau , \alpha \right) = \mathrm{Tr}\left(y^{W_0}q^{L_0}\right).
\end{equation}
As the contribution of the spin-3 black hole to $Z_{bh}$
is only known at large $c$ (i.e. the classical limit), a calculation of the character appearing on the right-hand
side of (\ref{holo reln}) to further orders in $c$ would give quantum corrections to the black hole partition function. This calculation was performed to leading order in $c$ in \cite{GutKraus} (c.f. the large central charge limit employed in \cite{GHJ}), but our results should hold for all $c$.


Finally, we note that as (\ref{holo reln}) gives a black hole partition function in terms of a CFT quantity, the thermodynamic properties of black holes can be investigated using CFT results - see e.g. \cite{DFP}. The
results we have obtained in this work may also find application in this
context, leading to new discoveries regarding the thermodynamic
properties of 2+1-dimensional black holes with spin-3 charge. 

\section*{Acknowledgements}
NJI would like to thank STFC for a doctoral training studentship. GMTW
would like to thank STFC for partial support under grant
ST/J002798/1. 
We would also like to the thank the referee for many
useful comments on the manuscript.

\end{document}